# Tiling large areas with luminescent solar concentrators: geometry effects

**ILYA SYCHUGOV**

*KTH-Royal Institute of Technology, Department of Applied Physics, 16440, Kista-Stockholm, Sweden*
*ilyas@kth.se*

**Abstract**

Luminescence solar concentrators act as semi-transparent photovoltaic cells and can be applied to large surface areas in modern urban environment. In this paper their optical efficiencies were analytically derived for different unit shapes as simple, integral-free expressions. Perimeter length increase for the same enclosed area requires more photodetectors, yet reduces optical losses per unit. An exact interplay between these two parameters is presented, which can be directly applied to the analysis of tiling for large areas. An explicit expression for the critical size of an LSC unit, above which its inner part becomes inactive, has been obtained. Fabrication of devices larger than specified by this limit should be discouraged from efficiency considerations.

## 1. Introduction

Solar light spatial concentration through fluorophore re-emission in a planar waveguide was introduced long ago to reduce the cost and to improve the performance of solar cells [1]. It was mainly based on organic dyes, embedded in a plastic slab. However, dye molecules suffer from strong re-absorption losses, which seriously limited practical application of such luminescent solar concentrators (LSCs) [2, 3]. Recent developments in quantum dots (QDs) make more favorable optical properties possible and thus bring the LSC concept closer to realization [4-7]. Covering large surfaces of, for example, facades of tall buildings with such semi-transparent solar cells may be considered as one of the goals [8, 9]. Here the exact geometrical shape and size of the device unit may play an important role in the total assembly efficiency.

Historically the "concentration ratio" has been considered as a main figure of merit for an LSC. By now the cost of solar cells dropped by about three orders of magnitude compared to the early days of LSC research. Therefore reducing the area covered by solar cells, and/or improving their performance by concentrating the incoming light flux may no longer be the primary motivation behind LSC development. Instead, possible modern applications include integration of such semi-transparent solar cells in urban environment [10]. Here the device efficiency of LSCs appears to be a more important parameter, even if it is associated with non-optimal solar cell performance or arrangement. From this application-driven perspective the incorporation of all loss mechanisms in the device efficiency description is a necessary condition for an adequate LSC analysis.

The analysis of optical losses in such planar waveguides is usually done by numerical integration [11] or by Monte-Carlo ray tracing [10, 12]. Indeed, simulations can follow evolution of nearly all aspects of the propagating light. In addition to its intensity, the spectrum, polarization, angle distribution, etc. can be tracked all the way to final distributions at the edges. In some cases, for example when only the total device efficiency needs to be estimated, such a complete and computationally demanding description may be redundant. We have previously shown that an analytical solution to the device efficiency can be obtained using only geometrical optics by probabilistic approach [13]. It was derived for a rectangular shape LSC as a function of the fluorophore and matrix parameters as well as device geometry.

Based on this analytical methodology here we turned to the question of optimal size and shape for an LSC unit in a large assembly. Intuitively one can associate two opposite effects with geometry of a single device. On the one hand, for a large area unit cell the total perimeter length, hence the overall number of needed photodetectors, will be lower. On the other hand, large devices will suffer from more optical losses per unit and, hence, reduced efficiency. Indeed, the LSC operation is based on the waveguiding effect, therefore light scattering, re-absorption and absorption in the matrix will become stronger. Here we quantitatively assessed the interplay between these two effects on examples of devices with common regular shapes and rectangle geometry. Based on the description including all the losses, we show that among regular shapes a hexagon has the most favorable geometry. While the square is not far away in performance, a substantial enhancement can be achieved by enclosing the same area with rectangles. It comes with a cost of a longer perimeter, requiring more photodetectors, and the exact relation between these two parameters is provided. We have also derived a single, integral-free formula for the device efficiency in case of a square shape LSC. More importantly, a simple criteria for the largest device size is introduced, which allows to quickly evaluate limitations of a particular matrix material for LSC units.

## 2. Shape Effect

We start our analysis by considering LSCs of regular shapes, enclosing the same area. As shown in Table 1, the perimeter length grows when shifting from a circle (radius $R$) to a triangle (side length $b$). In Table 1 the perimeter length is given in the units of a side length $a$ of the same area square, where $s$ is a hexagon side length.

First we analytically derive the optical path length $r$ distribution $p(r)$ for photons emanating from an isotropic emitter randomly located inside an LSC. It was done for various shapes and based on the algorithm used for a rectangle, as detailed in [13]. At this stage two dimensional geometry is considered and photons experience no attenuation. Results of analytical derivations were verified by the normalization condition: integrated $p(r)$ expressions over the whole range of optical paths should be equal to the enclosed area. Below the *normalized* (total integral is unity) probability density functions are presented. Thus, for a circle in a 2D case the result is

$$p_c(r) = \frac{\sqrt{4R^2 - r^2}}{\pi R^2}, \qquad 0 < r < 2R \tag{1}$$

For a hexagon the initial part is linear:

$$p_h(r) = \frac{4s - 2(1 - \pi\sqrt{3}/9)r}{s^2 \pi\sqrt{3}}, \qquad 0 < r < s \tag{2}$$

with more complex expressions for the remaining $s < r < 2s$ (presented in Appendix, A1). For a square the distribution is piece-wise:

$$p_s(r) = \begin{cases} \dfrac{4a - 2r}{\pi a^2}, & 0 < r < a \\ \dfrac{2r^2 - 4a\sqrt{r^2 - a^2}}{\pi a^2 r}, & a < r < a\sqrt{2} \end{cases} \tag{3}$$

And for a triangle the distribution is linear almost all the way:

$$p_t(r) = \frac{4\sqrt{3}b - (2\sqrt{3} + 4\pi/3)r}{\pi b^2}, \qquad 0 < r < b\sqrt{3}/2 \tag{4}$$

The exact expression for the remaining short range $b\sqrt{3}/2 < r < b$ is given in Appendix, A1.

Table 1. Perimeter lengths and average optical paths of photons in different shape LSC enclosing the same area

| Shape | Characteristic size, $a$ | Perimeter length, $a$ | Average optical path length, $a$ |
|---|---|---|---|
| Circle | $R = \pi^{-1/2} \approx 0.56$ | 3.55 | 0.479 |
| Hexagon | $s = \sqrt{2}/3^{3/4} \approx 0.62$ | 3.72 | 0.478 |
| Square | $a = 1$ | 4 | 0.473 |
| Triangle | $b = 2/3^{1/4} \approx 1.52$ | 4.56 | 0.461 |
| Rectangle 1 $\beta = 1.62$ | $w = \beta^{-1/2} \approx 0.79$ $h = \beta^{1/2} \approx 1.27$ | 4.12 | 0.467 |
| Rectangle 2 $\beta = 2.85$ | $w = \beta^{-1/2} \approx 0.59$ $h = \beta^{1/2} \approx 1.69$ | 4.56 | 0.444 |

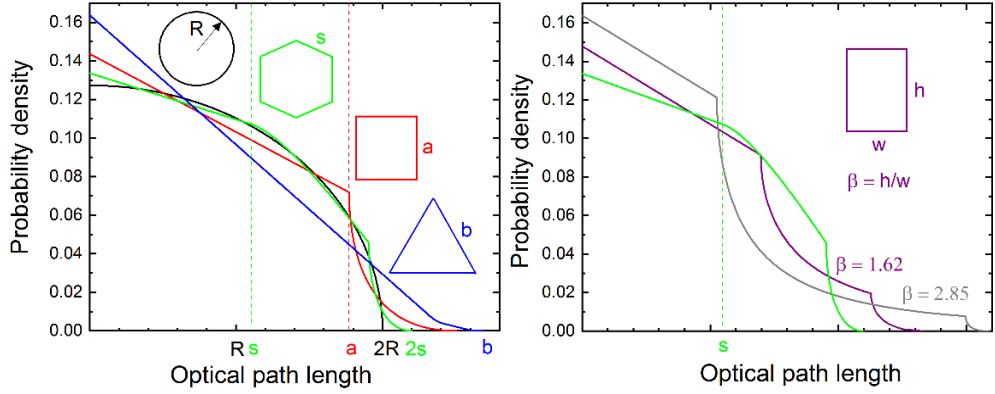

Figure 1. Normalized probability density functions for an emitted photon to travel a given optical path length in luminescent solar concentrators of different shapes enclosing the same area. (left) Regular shapes are compared. (right) A hexagon is compared with the same area rectangles of two different aspect ratios.

The derived distributions are shown in Figure 1, left, and their first moments (average optical paths) in Table 1. From Figure 1, left, it is seen how a longer perimeter (from circle to triangle) increases the fraction of shorter optical paths. It stems from a larger part of emitters located close to the light-collecting edge. At the same time, increasing the perimeter makes also longer paths possible for the photons travelling across the unit.

These two effects largely compensate each other, making the average optical path similar for all the considered regular shapes, as summarized in Table 1. So increasing the perimeter length by ~ 30%, when choosing a triangle instead of a circle, does not bring a meaningful increase in optical efficiency (the average optical path stays nearly the same). This is a first conclusion of this work, which may not be so obvious without explicit expressions (1)-(4) and Figure 1, left.

So, taking into account optical losses, the best *regular shape* for such a planar waveguide would obviously be a circle. Here the shortest perimeter is accompanied by nearly the same average optical path. Tiling without gaps, however, is not possible and flexible solar cells are required in this case to accommodate the circular edge. A more practical shape would be a hexagon, which can be tiled over large surfaces without gaps and does not require special photodetectors at the edges. A square does not differ substantially from a hexagon in this respect. On the other hand, the most inefficient way of enclosing an area in this case is a triangle. This geometry also suffers from a very non-uniform incoming light distribution along the edge, compared to other regular shapes [14].

Expanding the selection of unit shapes to *irregular* domain, one can immediately invoke a rectangle as an example, which is commonly used in modern large area facades. First, we need to introduce the rectangle aspect ratio $\beta$ as an auxiliary parameter. An explicit expression for $p_r(r)$ is provided in Appendix, A1. This normalized optical path probability distribution function for two values of $\beta$ is shown in Figure 1, right. It is seen that a larger fraction of shorter optical paths exists for rectangles, which results in the lower average value.

Quantitatively the average photon path $<r>$ in a rectangular slab can be analytically calculated as a first moment of the given optical path distribution. The result is a decaying function of $\beta$:

$$<r>_r = \frac{a\left(\beta^3 + 1 - (\beta^2 + 1)^{\frac{3}{2}} + 3\beta \ln(\sqrt{\beta^2 + 1} + \beta) + 3\beta^2 \ln\left(\frac{\sqrt{\beta^2 + 1} + 1}{\beta}\right)\right)}{3\pi\beta\sqrt{\beta}} \quad (5)$$

And the perimeter length of such a rectangle is

$$P = \frac{2(\beta + 1)}{\sqrt{\beta}} a \quad (6)$$

Numerical examples for two different rectangles are included in Table 1: the first corresponds to the well-known "golden ratio", and the second has the same perimeter length as the triangle. One can instantly notice that a rectangle with the same perimeter length as a triangle features lower optical losses. By increasing perimeter further (larger $\beta$), i.e. by deploying longer solar cells, the optical efficiency can be improved even more.

This effect is more clearly illustrated in Figure 2, where the two functions from Eq. (5) and Eq. (6) are shown. This graph shows how a longer perimeter helps to reduce an average optical path length for a rectangle, which is equivalent to reducing propagation losses in an LSC. The explicit effect on the total device efficiency, including all loss channels, will be shown later.

Before evaluating device efficiencies we note that for a 3D case the optical paths will be extended on average by a factor of $k \approx 1.14$ [13]. This is due to a narrow out-of-plane path distribution, which is limited by the escape cone. So 3D distributions for all the shapes can be represented as $q(l) \approx p(l/k)$. For example, in a circle the *normalized* probability density function becomes:

$$q_c(l) = \frac{\sqrt{4R^2 - (l/k)^2}}{\pi k R^2}, 0 < l < 2Rk \quad (7)$$

Explicit expression for the optical path distributions for other shapes in 3D are provided in the Appendix, A2.

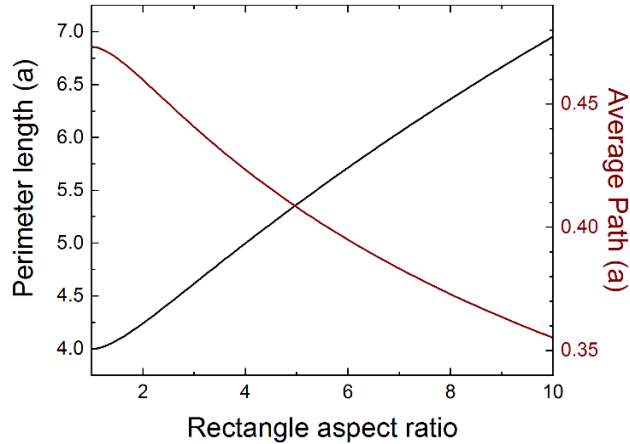

Figure 2 (brown) Shortening of the average optical path for isotropic emitters in different aspect ratio rectangles, enclosing the same area (from Eq. (5)). Units are the side length of the equivalent area square. (black) Related perimeter length increase is shown on the left axis (from Eq. (6)).

## 3. Critical size

Analytical expressions for $q(l)$ allow to obtain *explicitly* the fraction of emitted light reaching the edge of the device, i.e. the waveguiding or optical efficiency [%]:

$$f(x) = \int_0^{l_{max}} q(l) \exp(-xl) \, dl$$

where $l_{max}$ is the longest optical path in the system. Physically it corresponds to the optical efficiency when the only loss mechanism is the matrix absorption (with a loss parameter being the matrix absorption coefficient $x = \alpha$ [cm$^{-1}$]). For a circle ($l_{max} = 2R$) the integral yields:

$$f_c(\alpha) = -\frac{M_1(2\alpha Rk)}{\alpha Rk} \qquad (8)$$

where $M_1(\xi)$ is a modified Struve function of the second kind: $M_1(\xi) = L_1(\xi) - I_1(\xi)$, where $I_1(\xi)$ is a modified Bessel function, $L_1(\xi)$ is a modified Struve function (all functions are of the first order). Analytical expressions of $f(x)$ for other shapes are provided in the Appendix, A3. Formula (8) can be simplified for large arguments, using expansion of $M_1(\xi)$ (shown in the Appendix, A4) and leaving only the first term (a constant $-2/\pi$):

$$f_c(\alpha) \approx \frac{2}{\alpha R \pi k}, \qquad \alpha R \to \infty \qquad (9)$$

In Figure 3 these two expressions are shown in a log-log scale as a function of the dimensionless parameter $\alpha R$ (as a solid line for Eq. (8) and as a dotted line for Eq. (9)).

The transition from Eq. (8) to (9) signifies a critical size of the device. Indeed, in the latter regime only the emission close to the edge is collected. The emitted light from the central part is lost, so a device is no longer operating as a proper LSC. With increasing perimeter the active area close to the edge grows linearly, while the inactive middle part increases as a square of size. Hence the waveguiding efficiency becomes inversely dependent on the characteristic size, as Eq. (9) postulates. Therefore, in practice, it makes no sense to fabricate units larger than this critical size, which will render the inner part inactive.

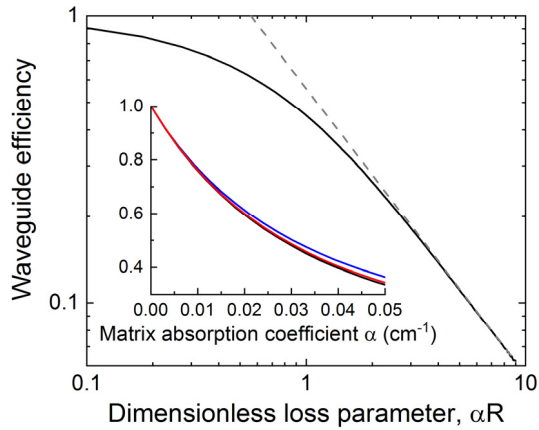

Figure 3. Dependence of the waveguiding efficiency of the dimensionless parameter αR. A dotted line is an asymptotic from Eq. (9) and the solid line is a full expression from Eq. (8). Inset shows waveguide efficiencies for a circle (black), a square (red), and a triangle (blue) for the same area units (R = 30 cm).

Numerically, the value of the critical dimensionless parameter can be extracted at the point in Fig. 3 where these two curves start converging. For $\alpha R = 1$ the difference in absolute efficiency is ~ 10%, while for $\alpha R = 2$ it is only ~ 1%. So for a circle one can approximately set the limit at $\alpha R_{cr} \leq 1.5$, where the difference is ~ 3%. Assuming the same optical path distribution (cf. Fig. 1), the criteria for a hexagon, enclosing the same area, is $\alpha \cdot s_{cr} \leq 1.7$, while for a square:

$$\alpha \cdot a_{cr} \leq 2.7 \qquad (10)$$

This is a second important conclusion of the presented analysis: a way to quickly estimate maximum useful LSC unit size is proposed. As a numerical example consider typical polymers, such as PMMA or OSTE with $\alpha \approx 0.03 - 0.04\ cm^{-1}$ as a matrix material [10, 15]. Then, according to Eq. (10), a square unit with a side length not more than ~ $70 - 90\ cm$ should be used. In further analysis we will limit considered device sizes by this criterion.

To illustrate the negligible effect of the regular shape choice on optical losses one can compare derived expressions $f(x)$ for LSCs enclosing the same area. Those are shown for a circle (black), a square (red) and a triangle (blue) in Figure 3, inset ($R = 30\ cm$). The x-axis is limited by the LSC critical parameter, as defined above. These functions essentially coincide for a small value of the loss parameter and deviate only by 1-2% for larger $\alpha$ in case of a triangle. So the considered here regular shapes indeed have very similar waveguiding efficiencies, as was also concluded from the average optical path comparison. This is despite a large difference in the perimeter length (cf. Table 1). Therefore we propose to use Eq. (8) as an approximation for all the shapes, except for the triangle. With this assumption simple expressions for the total LSC device efficiency can be derived for these geometries, including all the loss mechanisms.

The size limit for a rectangle can also be evaluated from the obtained criterion. Here we do not consider rectangles of the same area as regular shapes, but define the maximum reasonable aspect ratio for a given width. Note that it cannot be defined as clearly as for the case of a square, where the inner part becomes simply inactive. However, we can set the average optical path, corresponding to the critical size $a_{cr}$, as a critical optical path. Then the critical rectangle width [cm], corresponding to this average path, can be expressed as a function of the absorption coefficient $\alpha$ [cm$^{-1}$] and the aspect ratio $\beta$ from Eq. (5), Eq. (10) and Table 1:

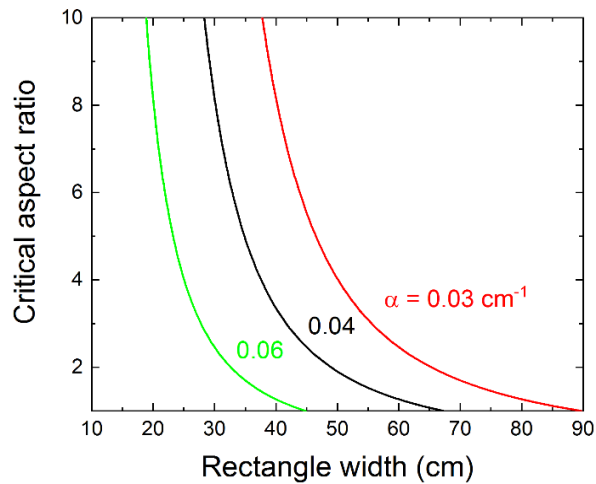

Figure 4. Critical aspect ratio of a rectangle LSC as a function of width for different values of the absorption coefficient.

$$w_{cr} \approx \frac{12\beta}{\alpha\left(\beta^3 + 1 - (\beta^2 + 1)^{\frac{3}{2}} + 3\beta \ln(\sqrt{\beta^2 + 1} + \beta) + 3\beta^2 \ln(\sqrt{\beta^2 + 1} + 1) - 3\beta^2 \ln(\beta)\right)}$$

Examples for some relevant matrix absorption coefficients are presented in Figure 4. It is shown instead as inverse function $\beta_{cr}(w)$, which indicates what the maximum aspect ratio can be tolerated for a given rectangle width.

## 4. Device optical efficiency

For more realistic description of the device performance other loss mechanisms need to be considered. Those include losses from scattering and re-absorption by the fluorophores. In general, the LSC optical efficiency $\gamma$ [%] in case of all possible losses can be evaluated using the waveguide efficiency function $f(x)$ using geometrical sum series as [13]:

$$\gamma = \frac{(1-T) \cdot \delta \cdot QY \cdot \eta \cdot f(\alpha_{sc} + \alpha_{re} + \alpha)}{1 - \frac{\delta \cdot \alpha_{sc} + \delta \cdot QY \cdot \alpha_{re}}{\alpha_{sc} + \alpha_{re} + \alpha}\left(1 - f(\alpha_{sc} + \alpha_{re} + \alpha)\right)}$$

Here $T$ is a transmitted fraction of the incoming sunlight ($1 - T$ is the absorbed fraction). Parameter $\eta = \epsilon_{PL}/\epsilon_{sun}$ is the energy conversion coefficient of the luminescence, defined as the ratio of luminescence and solar peak position energies. $QY$ is the quantum yield, and $\delta$ is a fraction of the emission to the waveguiding mode ($\delta = 75\%$ for $n = 1.5$ matrix material). The linear reabsorption coefficient is $\alpha_{re}$ [$cm^{-1}$], the linear scattering coefficient is $\alpha_{sc}$ [$cm^{-1}$], and the linear matrix absorption coefficient is $\alpha$ [$cm^{-1}$]. These should be taken for the wavelength of the fluorophore luminescence (corresponding to $\epsilon_{PL}$).

For a square shape, using Eq. (8), the resulting expression can be represented as a relatively simple formula (with $x = \alpha_{sc} + \alpha_{re} + \alpha$ as a loss parameter):

$$\gamma_s = \frac{(1-T) \cdot \delta \cdot QY \cdot \eta \cdot M_1(2xak/\sqrt{\pi}) \cdot x\sqrt{\pi}}{\delta(\alpha_{sc} + QY \cdot \alpha_{re})\left(akx + M_1(2xak/\sqrt{\pi}) \cdot \sqrt{\pi}\right) - akx^2} \quad (11)$$

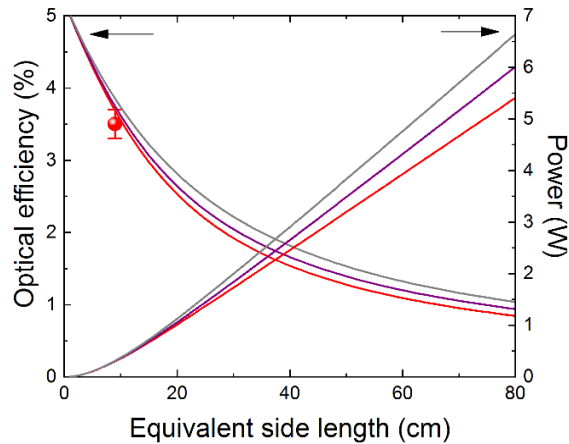

Figure 5. Device efficiency and corresponding optical power output for Si QD-based devices of the same area. Geometries are: a square (red), a rectangle with aspect ratio 1.62 (purple), a rectangle with aspect ratio 2.85 (grey). An experimental point is from measurements of a 9x9 cm² device.

As justified above it can also be used for hexagons with $a = 1.62 \cdot s$ and for circles with $a = R \cdot \sqrt{\pi}$. The formula (11) yields device efficiency for a square-shaped LSC without any integral representations, as contained in a more general formula for the rectangle [13]. We mark this as another important result of the presented analysis: a single integral-free formula for the efficiency of a square-shaped LSC was derived. Multiplying Eq. (11) by $\Phi a^2$ (solar flux $\Phi = 0.1\ W/cm^2$) gives optical efficiency in [W].

Unlike Eq. (8), where the only loss mechanism is matrix absorption, the total output here is not a function of a dimensionless parameter $xa$. The loss parameter $x$ and the waveguide geometry act as two independent parameters. So instead of a general criteria for establishing the critical size, the device performance for each particular material system should be evaluated. The criteria of Eq. (10) and Figure 4 then represent an upper limit.

To illustrate this result and to compare it with experiments we turn to Si QDs as fluorophores. They are often considered as one of possible candidates for this application due to a large Stokes shift and a good quantum yield [10, 15, 16]. In Figure 5 theoretical lines (red) of the LSC optical efficiency in [%] and in [W] are presented for a square (Eq. (11)). Two same area rectangles with the aspect ratios from Table 1 are also included (calculated from the general expression in [13]). The x-axis is a side length of the equivalent area square. The fluorophore parameters taken from the experiment are: QY = 55%, transparency for visible light T = 0.79 (a 3 mm thick film), re-absorption coefficient $\alpha_{re} = 0.007\ cm^{-1}$, scattering coefficient $\alpha_{sc} = 0.1\ cm^{-1}$, and matrix absorption coefficient $\alpha = 0.04\ cm^{-1}$ [17]. Although the device is not optimized in terms of scattering and has a high transmittance for visible light, we use it here mainly for comparison with the theory. Our experimental result for the estimated optical efficiency of ~ 3.4 % for a 9x9x0.3 cm³ piece (data point in Fig. 4) is indeed very close to the predicted values.

From Figure 5 it is seen how the efficiency improves with increasing the perimeter length by utilizing rectangles. In the beginning all the shapes have a similar efficiency. Here there are essentially no losses and almost all of the light is gathered from a given area. Then the larger fraction of shorter path lengths in rectangles (cf. Fig.1, b) starts playing a more important role, improving the efficiency. At this stage photons experiencing longer paths are preferentially removed from the system. In this way such shape-dependent quantitative analysis can be performed for any fluorophore/matrix/geometry LSC system.

In conclusion, we have derived analytical formulas for the optical path distribution for an isotropic emitter randomly placed in a planar waveguide of different shapes. Resulting expressions were used for the analysis of LSC performance as an interplay between the perimeter length and the propagation losses. Among regular shapes a hexagon has the best perimeter/loss combination. On other hand, at the expense of somewhat longer perimeter length, rectangular shapes with larger aspect ratios result in a better collection efficiency. This is because longer optical path lengths are suppressed, which, when optical losses are considered, results in a higher efficiency. Simple criteria for the critical unit size were obtained. The effect of the perimeter increase on the total device power output was quantitatively demonstrated for a given material system with a good match to the experiment.

**Funding.** Swedish Energy Agency (46360-1).

## Appendix

A1. Optical path length distribution in 2D for a hexagon with a side length $s$

$$p_h(r) = \begin{cases} \dfrac{4s - 2(1 - \pi\sqrt{3}/9)r}{s^2\pi\sqrt{3}}, & 0 < r < s \\[2mm] \dfrac{2s\sqrt{4r^2 - 3s^2}}{\sqrt{3}\pi r s^2} - \dfrac{2r}{3\pi s^2}\operatorname{asin}\left(\dfrac{s\sqrt{3}(2r^2 - 3s^2)\sqrt{4r^2 - 3s^2}}{2r^4}\right), & s < r < s\sqrt{3} \\[2mm] \dfrac{2r^2 - 8s\sqrt{r^2 - 3s^2}}{\sqrt{3}\pi r s^2} + \dfrac{2r}{3\pi s^2}\operatorname{asin}\left(\dfrac{\sqrt{3}(r^2 - 6s^2 + 2s\sqrt{r^2 - 3s^2})}{2r^2}\right), & s\sqrt{3} < r < 2s \end{cases}$$

For a rectangle with aspect ratio $\beta$ and width $w$ ($\beta > 1$)

$$p_r(r) = \begin{cases} \dfrac{2w + 2\beta w - 2r}{\pi\beta c^2}, & 0 < r < w \\[2mm] \dfrac{2r - 2\sqrt{r^2 - w^2}}{\pi c r}, & w < r < \beta w \\[2mm] \dfrac{2r^2 - 2\beta w\sqrt{r^2 - w^2} - 2c\sqrt{r^2 - \beta^2 w^2}}{\pi\beta w^2 r}, & \beta w < r < w\sqrt{\beta^2 + 1} \end{cases}$$

For a triangle with side length $b$

$$p_t(r) = \begin{cases} \dfrac{4\sqrt{3}b - (2\sqrt{3} + 4\pi/3)r}{\pi b^2}, & 0 < r < b\sqrt{3}/2 \\[2mm] u(r), & b\sqrt{3}/2 < r < b \end{cases}$$

$$u(r) = \dfrac{4\sqrt{3}}{\pi b^2}\left(b - \left(\dfrac{1}{2} - \dfrac{\pi\sqrt{3}}{18}\right)r - \dfrac{b\sqrt{4r^2 - 3b^2}}{2r}\right.$$

$$\left. + \dfrac{r\sqrt{3}}{6}\left(\operatorname{asin}\left(\dfrac{\sqrt{3}(\sqrt{4r^2 - 3b^2} - b)}{4r}\right) + \operatorname{asin}\left(\dfrac{\sqrt{3}(\sqrt{4r^2 - 3b^2} + b)}{4r}\right) - 2\operatorname{asin}\left(\dfrac{\sqrt{3}b}{2r}\right)\right)\right)$$

A2. For a 3D case in hexagon the properly normalized distribution function can be considered the same as in a circle for the same enclosed area: $R = 3^{3/4}s/\sqrt{2\pi} = c_0 \cdot s$, $c_0 \approx 0.91$ ($k \approx 1.14$):

$$q_h(l) \approx \dfrac{\sqrt{4c_0^2 s^2 - (l/k)^2}}{\pi k c_0^2 s^2}, \quad 0 < l < 2c_0 s k$$

For a triangle, considering only the linear part of the optical path distribution:

$$q_t(l) = \dfrac{4\sqrt{3}b - (2\sqrt{3} + 4\pi/3)(l/k)}{\pi k b^2}, \quad 0 < l < bk\sqrt{3}/2$$

For a square:

$$q_s(l) = \begin{cases} \dfrac{4a - 2(l/k)}{\pi k a^2}, & 0 < l < ka \\ \dfrac{2l^2/k - 4a\sqrt{l^2 - (ka)^2}}{\pi a^2 l k}, & ka < l < ka\sqrt{2} \end{cases}$$

A3. Waveguiding efficiency for a square:

$$f_s(x) = \frac{2\big(2axk - (ka\sqrt{2}x + 1) \cdot e^{-ka\sqrt{2}x} + 2e^{-kax} - 1\big)}{x^2 a^2 \pi k^2} - \frac{4}{a\pi k}\int_{ka}^{ka\sqrt{2}} \frac{\sqrt{l^2 - (ka)^2}}{l} \exp(-xl)\,dl$$

For a triangle:

$$f_t(x) = \frac{e^{-\frac{kb\sqrt{3}x}{2}}\big(6\sqrt{3} - 2\sqrt{3}bkx(6 - \pi) + 9bkx + 4\pi\big) + 12\sqrt{3}bkx - 4\pi - 6\sqrt{3}}{3x^2 b^2 \pi k^2}$$

For a rectangle (height $h$, width $w$, diagonal $d$):

$$f_r(x) = 2\left(\frac{(h+w)xk - (kdx + 1)e^{-kdx} + e^{-khx} + e^{-kwx} - 1}{x^2 hw\pi k^2} - \int_{kw}^{kd}\frac{\sqrt{l^2 - (kw)^2}}{w\pi k l}e^{-xl}\,dl \right.$$
$$\left. - \int_{kh}^{kd}\frac{\sqrt{l^2 - (kh)^2}}{h\pi k l}e^{-xl}\,dl\right)$$

A4. Expansion of the modified Struve function of the second kind (first order) for $\xi \to \infty$ (see, e.g. https://dlmf.nist.gov/11.6):

$$M_1(\xi) \approx \frac{1}{\pi}\sum_{k=0}^{\infty}(-1)^{k+1}\frac{\Gamma\left(k + \frac{1}{2}\right)}{\Gamma\left(\frac{3}{2} - k\right)}\left(\frac{2}{\xi}\right)^{2k} = -\frac{2}{\pi} + \frac{2}{\pi \xi^2} - \cdots$$